# Cross interface model for the thermal transport across interface between overlapped boron nitride nanoribbons

Wentao Feng[1,2], Xiaoxiang Yu[2], Yue Wang[2], Dengke Ma[2,3], Zhijia Sun[4], Chengcheng Deng[1,2]* and Nuo Yang[1,2]*

[1] State Key Laboratory of Coal Combustion, Huazhong University of Science and Technology, Wuhan 430074, P. R. China

[2] School of Energy and Power Engineering, Huazhong University of Science and Technology, Wuhan 430074, P. R. China

[3] NNU-SULI Thermal Energy Research Center (NSTER) & Center for Quantum Transport and Thermal Energy Science (CQTES), School of Physics and Technology, Nanjing Normal University, Nanjing 210023, P. R. China

[4] Institute of High Energy Physics, Chinese Academy of Sciences, Beijing 100049, China;

* Corresponding email: dengcc@hust.edu.cn (C.D.); nuo@hust.edu.cn (N.Y.)

**Abstract:**

The application of low-dimensional materials for heat dissipation requires a comprehensive understanding of the thermal transport at the cross interface, which widely exists in various composite materials and electronic devices. In this work, we proposed an analytical model, named as cross interface model (CIM), to accurately reveal the essential mechanism of the two-dimensional thermal transport at the cross interface. The applicability of CIM is validated through the comparison of the analytical results with molecular dynamics simulations for a typical cross interface of two overlapped boron nitride nanoribbons. Besides, it is figured out that the factor $\eta$ has important influence on the thermal transport besides the thermal resistance inside and between the materials, which is found to be determined by two dimensionless parameters from its expression. Our investigations deepen the understanding of the thermal transport at the cross interface and also facilitate to guide the applications of low-dimensional materials in thermal management.

# 1. Introduction:

The interfaces at the atomic scale play an important role in the thermal transport, which is essential for the performance of microelectronics, photonics, and thermoelectric devices [1, 2]. Thermal transport in materials could be greatly weakened by the existing interfaces [3-6]. Therefore, a deep understanding of interfacial thermal transport is crucial to improve the performance of various materials and devices for effective heat dissipation. Recently, the thermal transport across interface has been studied both experimentally and theoretically [7-11].

Due to their superior thermal conductivity, low-dimensional materials like carbon nanotube, graphene, boron nitride (BN) nanotube, BN nanoribbon et al, have elicited great interest as potential thermal interface materials [12-14]. They are usually used as fillers in composites or made into films [15-19]. However, the actual thermal conductivity of these composites and films are usually hard to meet the expectation, which implies that the thermal resistance between low-dimensional materials, like graphene-graphene interface and BN nanoribbon-BN nanoribbon interface, probably plays an important role in hindering thermal transport [20, 21]. So, it is necessary to deeply study the thermal transport across the interfaces between low-dimensional materials.

From the structure and morphology of composites or films observed in experiments, the low-dimensional materials, like graphene, CNT, BN, are staggered in parallel and formed many overlapped interfaces [3, 19, 22, 23]. And we called such kind of interface as cross interface. The schematic diagram of cross interface is shown in Fig. 1(a), and it is obvious that the thermal transport at the cross interface is a two-dimensional process. The heat is simultaneously transported inside and between the ribbons, which indicates its difference from the one-dimensional thermal transport of traditional interface [24-27]. However, it was generally treated as point contact or by approximation methods. For example, Zhong et al. reported molecular dynamics simulations on interfacial thermal resistance between carbon nanotubes with an overlapped interface. The interfacial thermal resistance was calculated by treating the overlapped region as a single planar interface between coaxial hot and cold nanotubes joined end to end [28]. In Yang et al.'s work, the total thermal resistance of the cross interface was treated as the sum of the contact thermal resistances and the two MWCNTs' thermal resistance with half of the overlapped length in series [29]. In Liu et al.'s work, the interfacial thermal conductance was calculated by the Fourier's formula, where the temperature difference was given by the average temperature difference between two graphene nanoribbons [30]. These previous works might introduce some approximations and lack the exploration for the essential mechanism to clearly describe the thermal transport at the cross interface. Therefore, there is a great demand for an accurate analytical model for thermal transport at the cross interface.

Our previous work derived a model for thermal transport at the cross interface between the same materials [31]. The interfacial thermal conductance between copper phthalocyanine nanoribbons was calculated by combining the model with experimental measurement. However, thermal transport problem at the interface between different materials tend to be more general [32-36]. And there needs fully and thoroughly understanding of the key influencing factors for the thermal transport at the cross interface.

In this work, we proposed an analytical model, named as cross interface model (CIM), for the thermal transport at the cross interface between different materials. We firstly deduce the analytical model to describe the two–dimensional thermal transport at the cross interface, and the expression of the total interfacial thermal resistance is demonstrated. Secondly, theoretical analysis is made to explore factor influencing thermal transport at the cross interface. Thirdly, an example of two overlapped BN nanoribbons (BNNRs) is constructed to compare the results of CIM with MD simulation results, which validates the accuracy of CIM. The influence of vacancy on the thermal transport is investigated by combining simulations with CIM.

## 2. Analytical Model

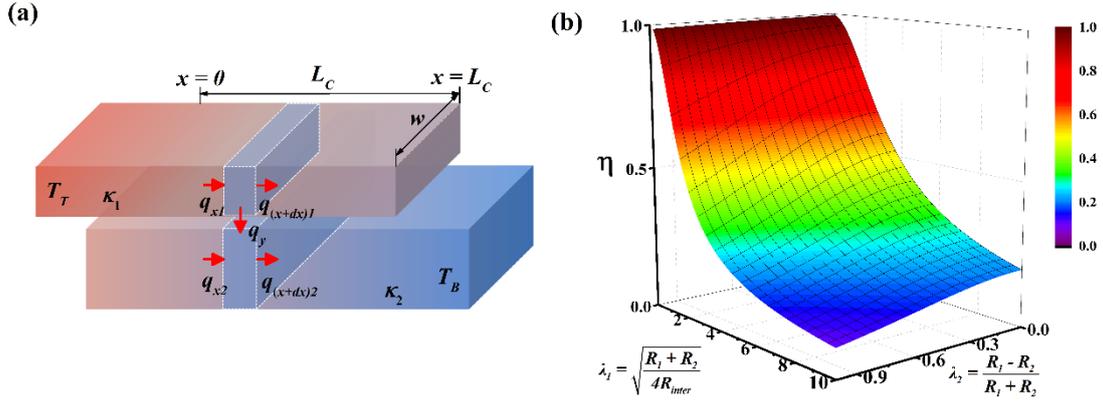

Figure 1. (a) Schematic diagram of cross interface model. (b) A factor $\eta$ in the expression of the total thermal resistance as a function of two influencing parameters, $\lambda_1$ and $\lambda_2$.

## 2.1 Deduction of the Cross Interface Model

In our work, an analytical model named cross interface model (CIM) is proposed and analytically deduced to investigate the thermal transport at the cross interface. As shown in Fig. 1(a), cross interface is schematically represented by two ribbons with an overlapped region. Here, the two ribbons are assumed with different thermal conductivity ($\kappa_1$ and $\kappa_2$) and cross section area ($A_1$ and $A_2$) for universal applicability. Heat source and sink are imposed on the end of top and bottom ribbon respectively. In the steady state, the heat conduction can be described based on the Fourier' law and energy conservation.

At the overlapped region, the heat conduction equations for each ribbon are given as below.

$$\kappa_1 \frac{d^2 T_T}{dx^2} A_1 - G_{CA}(T_T - T_B) w = 0, \quad 0 < x < L_C \tag{1a}$$

$$\kappa_2 \frac{d^2 T_B}{dx^2} A_2 + G_{CA}(T_T - T_B) w = 0, \quad 0 < x < L_C \tag{1b}$$

where $\kappa_1$ and $\kappa_2$ are the thermal conductivity of the top and bottom ribbon; $G_{CA}$ is the interfacial thermal conductance per unit area; and $T_T$ and $T_B$ are the temperature of the top and bottom ribbons, respectively; $w$ is the width of the ribbion, and $L_C$ is the overlapped length. We assume that all the thermal properties are constant in the ribbon. So the CIM is not suitable when temperature excursions of either ribbon are large enough to invalidate this assumption. These equations are similar to the two-temperature model and the two-channel heat transport model [37, 38]. By combining Eq. (1a) and (1b), the following equations can be obtained:

$$\frac{d^2(T_T - T_B)}{dx^2} = \gamma_1^2 (T_T - T_B), \quad 0 < x < L_C \tag{2a}$$

$$\frac{d^2(T_T + T_B)}{dx^2} = \gamma_2 (T_T - T_B), \quad 0 < x < L_C \tag{2b}$$

where $\gamma_1 = \sqrt{G_{CA} w / \kappa_1 A_1 + G_{CA} w / \kappa_2 A_2}$, $\gamma_2 = G_{CA} w / \kappa_1 A_1 - G_{CA} w / \kappa_2 A_2$.

Boundary conditions are needed to determine the solutions. The temperature of top ribbon at the left edge of the interface ($T_H$) and the temperature of bottom ribbon at the right edge of the interface ($T_S$) can be obtained. Moreover, the heat dissipation through the end of ribbon is negligible. Thus, two sets of boundary conditions are given by

$$\frac{dT_T}{dx}\bigg|_{x=L_C} = 0, \quad T_T\big|_{x=0} = T_H \tag{4a}$$

$$\frac{dT_B}{dx}\bigg|_{x=0} = 0, \quad T_B\big|_{x=L_C} = T_S \tag{4b}$$

Then we can solve the temperature distribution functions as

$$T_T = \frac{1}{2}\left\{\left(\frac{\gamma_2}{\gamma_1^2} + 1\right)\left[a \times e^{-\gamma_1 x} + b \times e^{\gamma_1 x}\right] + cx + d\right\} \tag{5a}$$

$$T_B = \frac{1}{2}\left\{\left(\frac{\gamma_2}{\gamma_1^2} - 1\right)\left[a \times e^{-\gamma_1 x} + b \times e^{\gamma_1 x}\right] + cx + d\right\} \tag{5b}$$

where

$$a = \frac{2(T_S - T_H)\gamma_1^2 \left[(\gamma_2 - \gamma_1^2) - e^{\gamma_1 L_C}(\gamma_2 + \gamma_1^2)\right]}{\gamma_1 L_C (\gamma_2^2 - \gamma_1^4)(e^{-\gamma_1 L_C} - e^{\gamma_1 L_C}) + 2(\gamma_2^2 + \gamma_1^4)(e^{-\gamma_1 L_C} + e^{\gamma_1 L_C}) - 4(\gamma_2^2 - \gamma_1^4)} \tag{6a}$$

$$b = a \times \frac{(\gamma_2 - \gamma_1^2) - e^{-\gamma_1 L_C} \times (\gamma_2 + \gamma_1^2)}{(\gamma_2 - \gamma_1^2) - e^{\gamma_1 L_C} \times (\gamma_2 + \gamma_1^2)} = a \times B \tag{6b}$$

$$c = a \times \frac{(\gamma_2^2 - \gamma_1^4)(e^{-\gamma_1 L_C} - e^{\gamma_1 L_C})}{\gamma_1 \left[(\gamma_2 - \gamma_1^2) - e^{\gamma_1 L_C}(\gamma_2 + \gamma_1^2)\right]} = a \times C \tag{6c}$$

$$d = 2T_H - a \times \frac{2(\gamma_2^2 - \gamma_1^4) - (\gamma_2 + \gamma_1^2)^2 (e^{-\gamma_1 L_C} + e^{\gamma_1 L_C})}{\gamma_1^2 \left[ (\gamma_2 - \gamma_1^2) - e^{\gamma_1 L_C}(\gamma_2 + \gamma_1^2) \right]} \quad (6d)$$

Note that all the heat will flow through the contact region between the two ribbons, thus the heat energy can be calculated by integrating the heat flux over the contact area,

$$J = \int_0^{L_C} \theta_1 G_{CA} w \, dx = \frac{-G_{CA} w a}{\gamma_1} \left[ e^{-\gamma_1 L_C} - 1 - B(e^{\gamma_1 L_C} - 1) \right] \quad (7a)$$

By simplifying Eq. (7a), we can get

$$J = G_{CA} w \int_0^{L_C} (T_T - T_B) dx = G_{CA} w \overline{(T_T - T_B)} \quad (7b)$$

where $\overline{(T_T - T_B)}$ is the average temperature difference at the cross interface. It means that the interfacial thermal conductance can be accurately calculated if we know the heat flux and average temperature difference.

Thus, the main analytical result, the expression of the total thermal resistance ($R_{total}$), is obtained as

$$\begin{aligned} R_{total} &= \frac{T_T|_{x=0} - T_B|_{x=L_C}}{J} = \frac{L_C}{\kappa_1 A_1 + \kappa_2 A_2} + \frac{1}{\eta} \times \frac{1}{G_{CA} w L_C} \\ &= \frac{R_1 R_2}{R_1 + R_2} + \frac{1}{\eta} \times R_{inter} = R_{intra} + \frac{1}{\eta} \times R_{inter} \end{aligned} \quad (8)$$

where $R_1$ and $R_2$ are the original thermal resistance of top and bottom ribbon respectively, which are called as the in-plane thermal resistance. $R_{intra}$ denotes the intra-ribbon thermal resistance assuming the in-plane thermal resistance of two nanoribbons obey the parallel law, and $R_{inter}$ denotes the inter-ribbon thermal resistance, which is the inverse of interfacial thermal conductance. The factor, $\eta$ ($\eta < 1$), is expressed as

$$\eta = \frac{1}{\frac{\gamma_1 L_C}{2} \left( \frac{\gamma_2^2}{\gamma_1^4} \tanh \frac{\gamma_1 L_C}{2} + \coth \frac{\gamma_1 L_C}{2} \right)} = \frac{1}{\lambda_1 (\lambda_2 \tanh \lambda_1 + \coth \lambda_1)} \quad (9)$$

The physical meanings of $\lambda_1$ and $\lambda_2$ are discussed in the following section.

## 2.2 Discussion of Cross Interface Model

As mentioned before, the cross interface structures widely exist in composites and play an essential role in heat dissipation. So, it is quite important to analyze and improve the thermal transport through the cross interface. It is noticed in the expression of the total thermal resistance, Eq. (8), that the factor $\eta$ plays an important role in the optimization of thermal transport at the cross interface. In other words, in order to improve the thermal transport, we not only need to reduce the intra-ribbon and inter-ribbon thermal resistance ($R_{intra}$ and $R_{inter}$), but also need to increase the factor $\eta$. Here we address this issue and give further discussion on the factor $\eta$.

According to Eq. (9), $\eta$ is related to two dimensionless parameters $\lambda_1 = \gamma_1 L_C/2$, and $\lambda_2 = \gamma_2/\gamma_1^2$. The definition of $\gamma_1$ and $\gamma_2$ has been given before that $\gamma_1 = \sqrt{G_{CA}w/\kappa_1 A_1 + G_{CA}w/\kappa_2 A_2}$ and $\gamma_2 = G_{CA}w/\kappa_1 A_1 - G_{CA}w/\kappa_2 A_2$. Thus, the following expressions can be obtained through some transformation:

$$\lambda_1 = \frac{\gamma_1 L_C}{2} = \frac{1}{2}\sqrt{\left(\frac{L_C}{\kappa_1 A_1} + \frac{L_C}{\kappa_2 A_2}\right)\bigg/\frac{1}{G_{CA}wL_C}} = \sqrt{\frac{R_1 + R_2}{4R_{inter}}} \tag{10a}$$

$$\lambda_2 = \frac{\gamma_2}{\gamma_1^2} = \left(\frac{L_C}{\kappa_1 A_1} - \frac{L_C}{\kappa_2 A_2}\right)\bigg/\left(\frac{L_C}{\kappa_1 A_1} + \frac{L_C}{\kappa_2 A_2}\right) = \frac{R_1 - R_2}{R_1 + R_2} \tag{10b}$$

According to the Eq. (10a,b), the physical meaning of $\lambda_1$ and $\lambda_2$ can be revealed. $\lambda_1$ can represent the ratio of in-plane thermal resistance to inter-ribbon resistance, and $\lambda_2$ can evaluate the difference of in-plane thermal resistance between the two ribbons.

Here, the influence of $\lambda_1$ and $\lambda_2$ on factor $\eta$ is discussed. Based on the above expression, the relationship of $\eta$ as a function of $\lambda_1$ and $\lambda_2$ is shown in Fig. 1(b). It can be seen that, when $\lambda_1$ approaches 0, that is, when the in-plane thermal resisitance ($R_1+R_2$) is much larger than inter-ribbon one ($R_{inter}$), $\eta$ approaches 1. And it is worth noting that $\eta$ decreases sharply with $\lambda_1$ and then gradually levels off. When $\lambda_1$ is equal to 10, the value of $\eta$ is only 0.1. Meanwhile, $\eta$ decreases much more slightly with $\lambda_2$ when compared with $\lambda_1$, that means, large difference between the thermal conductivity of two ribbons could only slightly influence $\eta$. In a word, increasing $\eta$ is an effective way for the optimization of thermal transport at the cross interface. The analysis indicates that, even if the sum of intra-ribbon and inter-ribbon resistance ($R_{intra}+R_{inter}$) are fixed, the thermal transport could still be improved by decreasing the in-plane thermal resistance rather than the inter-ribbon thermal resistance. In addition, reducing the difference between in-plane thermal resistances of two ribbons could also help to slightly enhance the thermal transport.

In this part, the proposed CIM is derived and anlyzed to give a deep understanding of the thermal transport at the cross interface. And the factor $\eta$ is figured out from the expression of the total thermal resistance, whose increase could improve the thermal transport. More importantly, the parameters that influence $\eta$ are further discussed.

In order to verify the feasibility of CIM under different conditions and confirm the influence of parameters $\lambda_1$, $\lambda_2$ on the factor $\eta$, a typical example of two overlapped boron nitride nanoribbons (BNNRs) is given below by molecular dynamics (MD) simulations to compare with the analytical model. Different conditions are conducted by changing thermal transport properties at the cross interface. It is well-known that the thermal transport inside and between two nanoribbons could be modulated by many strategies, such as vacancies, covalent bonds, isotope, et al [39-43]. We consider the impact of vacancy, which is a kind of widely studied defect in nanoribbons [44-47]. And the systems of two BNNRs with different total vacancy concentration $\rho_{total}$ and different vacancy concentration ratio between two nanoribbons $\rho_{top}/\rho_{bottom}$ are constructed in the MD simulations.

## 3. MD Simulation Details

The thermal transport at the cross interface of BN nanoribbons is numerically calculated by means of nonequilibrium molecular dynamics (NEMD), which is a kind of useful technique to study thermal properties of materials [48-50]. The structure consists of a pair of BN nanoribbons, each of them is 27.33 nm long and 4.12 nm wide. Some vancancies are introduced to the BN nanoribbons periodically as shown in Fig. 2(a). The fixed boundary condition is used along z direction. The optimized Tersoff potential is applied to describe covalent bonding in BN nanoribbons, which has successfully reproduced the thermal transport properties of BN before [51]. The detailed parameters of optimized Tersoff potential could be found in Supplementary

Material Table S1. The interactions between BN nanoribbons are van der Waals forces modeled by the Lennard-Jones potential, $V_{ij} = 4\varepsilon\left[(\sigma/r)^{12} - (\sigma/r)^{6}\right]$, with the parameters calculated from the universal force field (UFF): $\varepsilon_{B-N}$ = 4.833 meV, $\sigma_{B-N}$ = 3.449 Å, $\varepsilon_{B-B}$ = 7.806 meV, $\sigma_{B-B}$ = 3.638 Å, $\varepsilon_{N-N}$ = 2.992 meV, $\sigma_{N-N}$ = 3.261 Å [52]. The cutoff distance is set as 8.5 Å. Here $r$ is the distance between two atoms. Verlet algorithm is adopted to integrate the discrete differential equations of motions. The time step is set as 0.25 fs.

We relax the BN nanoribbon structure in the canonical ensemble (NVT) and microcanonical ensemble (NVE). After relaxation, the heat source with a higher temperature 320 K is applied to the atoms in red region and the heat sink with a lower temperature 280 K is applied to atoms in blue region. Then simulations are performed for 2.5 ns to reach a steady state. After that, a time average of the temperature and heat current is performed for 5 ns to get the temperature profile and the value of heat flux. The process of NEMD simulation could also be found in Supplementary Material Table S1. All of our simulations are performed by large-scale atomic/molecular massively parallel simulator (LAMMPS) packages [53]. The total thermal resistance ($R_{total}$) is calculated by the ratio of heat flux to the total temperature difference $R_{total} = \Delta T_{total} / J$.

## 4. Simulation Results and Discussions

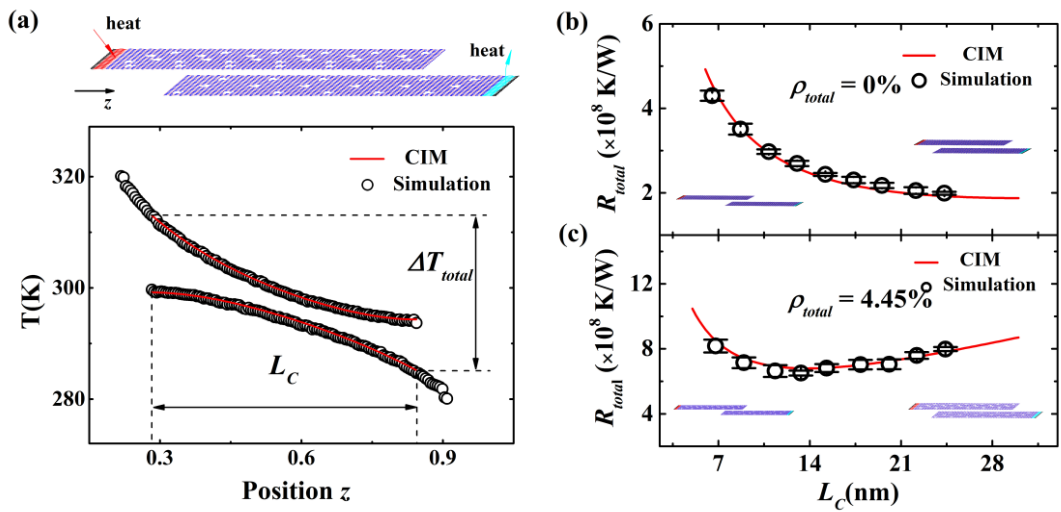

Figure 2. (a) Schematic of the BN nanoribbon structure and temperature profile. The red region and the blue region in the structure represent the heat source and heat sink separately. The expression of the temperature is fitted to the simulation result as the red line. (b,c)The overlapped length dependence of total thermal resistance with different total vacancy concentration: (b) total vacancy concentration=0; (c) total vacancy concentration=4.45%. The red line is the predicted curve of CIM.

Fig. 2 shows an example of NEMD calculation for the thermal transport of BNNRs cross interface. A typical temperature profile of the interface structure with 4.45% total vacancy concentration is presented in Fig. 2(a). In order to preliminarily verify the accuracy of CIM, the derived expression of temperature profile in CIM, Eq. (5), is fitted to NEMD calculation results. Here, the coefficient of determination, $R^2$, is used to evaluate the quality of the fitting [54]. When the value of $R^2$ equals to 1, it indicates that the model perfectly fits the data. Among our fitting results of all tested structures, whose total vacancy concentration range from 0 to 7.5%, high values of $R^2$ (above 0.998) are obtained. It suggests that the proposed CIM can accurately describe the thermal transport at the cross interface. Moreover, the thermal conductivity $\kappa$ and interfacial thermal conductance $G_{CA}$ could be get in the fitting process.

To further validate the accuracy of CIM, the total interfacial thermal conductance ($R_{total}$) calculated by CIM is compared with the corresponding simulation results in Fig. 2(b,c). In our simulation, the length of each BN nanoribbon is kept constant while the overlapped length of the cross interface $L_C$ is changed. Also, in order to test the performance under different conditions, some periodic vacancies are introduced to BNNRs and the results of two different structures, (b) with no vacancy and (c) with total vacancy concentration of 4.45% can be compared.

The calculation of the total interfacial thermal conductance ($R_{total}$) by CIM could be easily realized if we know the thermal conductivity of BNNRs ($\kappa_1$, $\kappa_2$) and interfacial thermal conductance ($G_{CA}$) according to its expression. Therefore, we firstly calculate $\kappa_1$, $\kappa_2$ and $G_{CA}$ by MD simulation respectively and then plug them into the expression of $R_{total}$, Eq. (8). Thus, the model calculation values of $R_{total}$ could be obtained as the red line in Fig. 2(b,c). As for NEMD calculations, five simulations are performed with different initial states for one structure, and the average value is shown, where the error bar is the standard deviation.

In Fig. 2(b,c), the total thermal resistance ($R_{total}$) of CIM calculation matches well with the simulation values for different BNNRs structures. The results verify that the proposed cross interface model is applicable. Besides, it is found that $R_{total}$ decreases initially and then increases with overlapped length ($L_C$). The reason is that with the growth of $L_C$, in-plane thermal resistance ($R_1$ and $R_2$) is enlarged but interfacial thermal resistance ($R_{inter}$) is reduced. Meanwhile, the increase of $R_1$ and $R_2$ with $L_C$ is linear while the reduction of $R_{inter}$ would gradually slow down according to their expression, which comprehensively caused the varying trend of $R_{total}$.

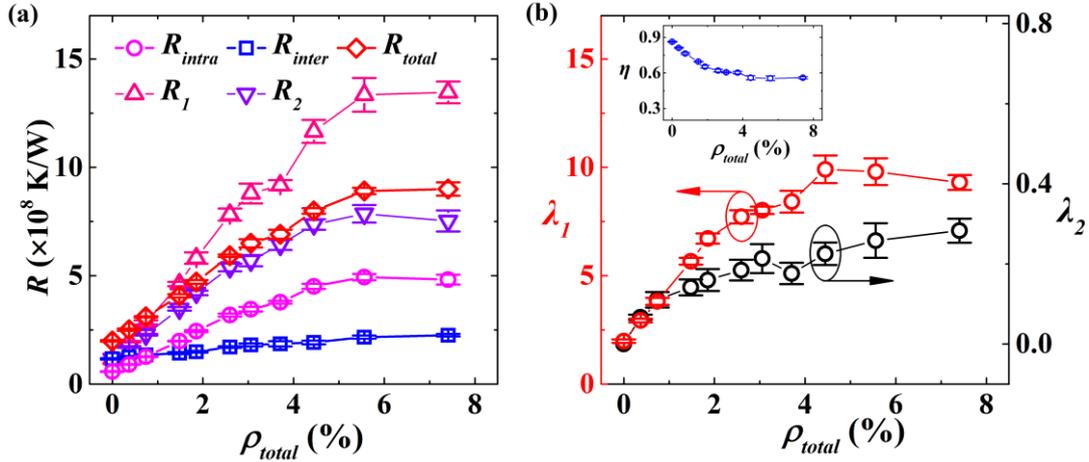

Figure 3. The effect of total vacancy concentration $\rho_{total}$ on thermal transport at the cross interface. The vacancy concentration ratio between two nanoribbons $\rho_{top}/\rho_{bottom}$ is set to 5:3. (a) The total thermal resistance ($R_{total}$), thermal resistance in ($R_1$, $R_2$ and $R_{intra}$) and between ($R_{inter}$) BNNRs. (b) Two dimensionless parameters $\lambda_1$ and $\lambda_2$ versus total vacancy concentration $\rho_{total}$. The inset shows the change of factor $\eta$ with total vacancy concentration.

Then the influence of vacancy concentration $\rho$ on thermal transport at the cross interface is studied through CIM as depicted in Fig. 3. In the simulations, vacancy concentration ratio between two ribbons $\rho_{top}/\rho_{bottom}$ is set to 5:3. Fig. 3(a) indicates that the total thermal resistance ($R_{total}$) increases obviously with vacancy concentration. It could be observed from the expression of $R_{total}$ that the thermal transport at the cross interface is only related to three parameters, including the intra-ribbon thermal resistance ($R_{intra}$), inter-ribbon thermal resistance ($R_{inter}$) and factor $\eta$. We then calculate them and explore the reasons for their changes.

The increase of $R_{intra}$ shown in Fig. 3(a) stems from the enlarged in-plane thermal resistance ($R_1$ and $R_2$). Previous studies also proved that the thermal transport in BNNRs would be greatly deteriorated by vacancies because of the strengthened phonon scattering [55]. As for the increase of $R_{inter}$, it is mainly caused by the decrease of the adhesion energy between two BNNRs with $\rho_{total}$, which is confirmed in Supplementary Material Fig. S1(a) [56-58].

The inset in Fig. 3(b) indicates that $\eta$ reduces with vacancy concentration and finally reaches a plateau. As shown in Fig. 1(b), the factor $\eta$ would decrease with the growth of $\lambda_1$ and $\lambda_2$. $\lambda_1$ is the ratio of the in-plane thermal resistances ($R_1$, $R_2$) to inter-ribbon thermal resistance ($R_{inter}$) and $\lambda_2$ evaluates the difference of in-plane thermal resistance between two ribbons. It is displayed in Fig. 3(a) that the increase of $R_1$ and $R_2$ with vacancy concentration are much more obvious than that of $R_{inter}$, which leads to the increase of $\lambda_1$ shown in Fig. 3(b). Also, the difference between two in-plane thermal resistance also increase according to Fig. 3(a) and the enlarged difference finally leads to the increase of $\lambda_2$. Therefore, it is reasonable that with the introduction of vacancies, $\eta$ would gradually decrease. Besides, when more vacancies are introduced to BNNRs, the enhancement of $R_1$, $R_2$ would slow down, which could explain the convergence of $\eta$. In a word, the change of $R_{intra}$, $R_{inter}$ and $\eta$ contribute to the increase of $R_{total}$ jointly.

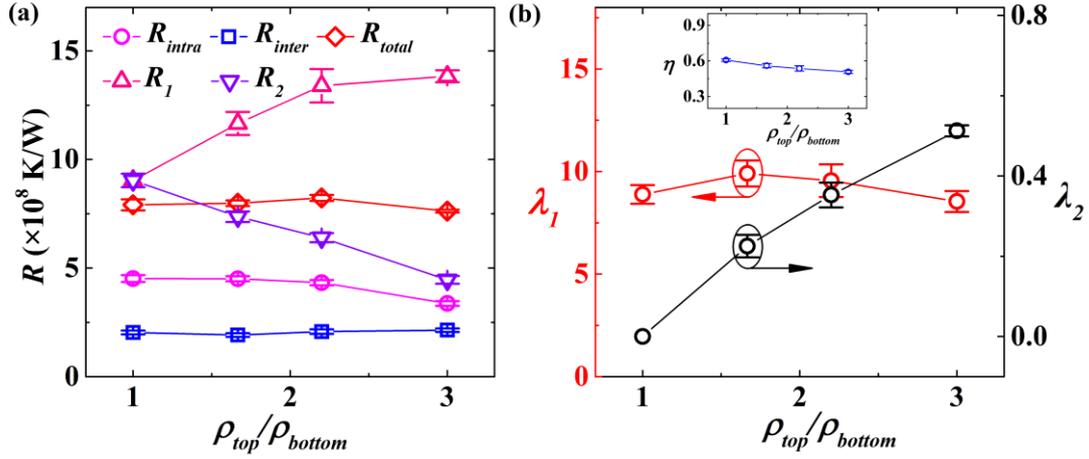

Figure 4. The effect of vacancy concentration ratio between two ribbons $\rho_{top}/\rho_{bottom}$ on thermal transport at the cross interface. The total vacancy concentration $\rho_{total}$ is set to 4.45%. (a) The total thermal resistance ($R_{total}$), thermal resistance in ($R_1$, $R_2$ and $R_{intra}$) and between ($R_{inter}$) BNNRs. (b) Two dimensionless parameters $\lambda_1$ and $\lambda_2$ versus vacancy concentration ratio. The inset shows the change of factor $\eta$ with vacancy concentration ratio.

In Fig. 3(b), the introduction of vacancies causes the increase of $\lambda_1$ and $\lambda_2$ simultaneously, but their contributions to the thermal transport at the cross interfafce couldn't be seperated. To quantify the impact of $\lambda_2$ separately, the total vacancy concentration $\rho_{total}$ is kept constant at 4.45%, while the ratio between two ribbons ($\rho_{top}/\rho_{bottom}$) is changed so as to change $\lambda_2$ only. In Fig. 4(b), it illustrates that $\lambda_1$ is not sensitive to the vacancy concentration ratio while $\lambda_2$ becomes more sensitive. It results from the fact that vacancy concentration in the top ribbon increases while the other decreases with vacancy concentration ratio, which enlarges $R_1$ but reduces $R_2$ and keeps $R_{inter}$ constant as shown in Fig. 4(a). The change of $R_1$ and $R_2$ are because of the strengthened and weakened phonon scaterring in the top and bottom BNNR repectively. Meanwhile, the constant adhesion energy shown in Figure S1(b) explains the unchanged $R_{inter}$. Comparing Fig. 4(b) with Fig. 3(b), the influence of vacancy concentration ratio on $\lambda_2$ is more obvious than that of total vacancy concentration. Even so, $\eta$ decrease much more slightly as can be seen in the inset of Fig. 4(b). Thus, it is further confirmed that the decrease of $\eta$ with total vacancy concentration is mainly due to the change of $\lambda_1$.

## 5. Conclusion

In summary, we propose an analytical model, CIM, to accurately reveal the essential mechanism of thermal transport at the cross interface. Compared with previous method, CIM considers two-dimensional instead of one-dimensional thermal transport at the cross interface. The merit of this work lies in the fact that we not only deduce the analytical model, but also validate its accuracy by comparing the analytical results with NEMD simulation of overlapped BNNRs under different conditions. Furthermore, it is figured out from the analytical model that increasing the dimensionless factor $\eta$ could also improve the thermal transport at the cross interface. Through analytical analysis, the factor $\eta$ could be enlarged by decreasing the ratio of in-plane to inter-ribbon thermal resistance or the difference between in-plane thermal resistances of two ribbons. And it is further confirmed by changing the total vacancy concentration and vacancy concentration ratio of two overlapped BNNRs in MD simulations. Our studies provide a new analytical model to deepen the understanding of the thermal transport at the cross interface and also explore effective methods to improve the heat dissipation of low-dimensional materials in practical applications.

## Supplementary Material

See supplementary material for detailed simulation settings and the adhesion energy between two BNNRs.

## Acknowledgement

This work was financially supported by the National Natural Science Foundation of China (No. 51606072 (C. D.), No. 51576076 (N. Y.), No. 51711540031 (N. Y.)), the

Natural Science Foundation of Hubei Province (No. 2017CFA046 (N. Y.)) and the Fundamental Research Funds for the Central Universities, HUST (No. 2019kfyRCPY045). We are grateful to Xiao Wan, Han Meng and Rulei Guo for useful discussions. The authors thank the National Supercomputing Center in Tianjin (NSCC-TJ) and High Performance Computer Cluster in Huazhong University of Science and Technology (HUST-HPCC) for providing assistance in computations.